\documentclass{aa}
\usepackage{graphicx}
\usepackage[]{natbib}
\def\lunits{$\rm erg\,s^{-1}$~}
\def\funits{$\rm erg\,cm^{-2}\,s^{-1}$~}
\def\cunits{$\rm cm^{-2}~$}
\def\lxlo{$\rm L_X-L_{[OIII]}$}

\def\xmm{{\it XMM-Newton~}}

\begin{document}
 \title{Comparison between the Luminosity functions of X-ray and [OIII] selected AGN}
  \titlerunning{Optical vs. X-ray AGN selection}
    \authorrunning{I. Georgantopoulos \& A. Akylas}
   \author{I. Georgantopoulos\inst{1},
           A. Akylas \inst{1},
            }
   \offprints{I. Georgantopoulos, \email{ig@astro.noa.gr}}
   \institute{Institute of Astronomy \& Astrophysics,
              National Observatory of Athens, 
 	      Palaia Penteli, 15236, Athens, Greece \\
              }
   \date{Received ; accepted }

\abstract{We investigate claims according to which the X-ray selection of AGN is not as efficient 
compared to that based on  [OIII] selection because of the effects of X-ray absorption. 
We construct the predicted X-ray luminosity function both for all Seyferts as well as separately 
for Seyfert-1 and Seyfert-2 type galaxies, by combining the optical AGN [OIII] luminosity functions 
derived in SDSS with the corresponding $\rm L_X-L_{[OIII]}$ relations.
These relations are derived from   {\it XMM-Newton} observations of all Seyfert galaxies 
in the Palomar spectroscopic sample of nearby galaxies after correction for X-ray absorption and optical reddening.  
We compare the predicted X-ray  luminosity functions with those  actually observed in the local Universe by 
{\it HEAO-1}, {\it RXTE} as well as {\it INTEGRAL}. The last luminosity function is derived in the 17-60 keV 
region and thus is not affected by absorption even in the case of Compton-thick sources.   
In the common luminosity regions, the optically and X-ray selected  Seyfert galaxies show reasonable agreement. 
We thus find no  evidence  that the [OIII] selection provides a more robust tracer of powerful AGN compared 
to the X-ray. Still, the optical selection probes less luminous Seyferts compared to the current X-ray surveys. 
These low luminosity levels, are populated by a large number of X-ray unobscured Seyfert-2 galaxies.     
}
\maketitle
%
%________________________________________________________________

\section{Introduction} 
Since the detection of the nearby AGN 3C273 in X-rays by {\it UHURU} (Kellogg et al. 1971), 
X-ray observations have been considered to be the primary tool for selecting AGN. 
Recently, the deepest ever observations  in 
the Chandra Deep Field North and South (Alexander et al. 2003, Giaconni et al. 2002, 
Luo et al. 2008) have resolved 80-90\% of the extragalactic 
X-ray light, the X-ray background, in the 2-10 keV band. These observations reveal a sky density of about 5000 
sources per square degree (Bauer et al. 2004), the vast majority of which are AGN (for a review see Brandt 
\&  Hasinger 2005). These surveys allowed the derivation of the luminosity function and probed with 
good accuracy the accretion history of the Universe 
(Ueda et al. 2003, La Franca et al. 2005, Barger et al. 2005).

In the local Universe, the X-ray luminosity function has been derived from the  wide-angle surveys 
of {\it RXTE} and {\it HEAO-1}. Sazonov \& Revnitsev (2004) derived the luminosity function for bright AGN, 
detected at fluxes $>2.5\times 10^{-11}$ \funits in the 3-20 keV band, in the {\it RXTE} slew survey.
Shinozaki et al. (2006) derived the luminosity function of AGN  using the all-sky HEAO-1 in the 2-10 keV band. 
Even these hard X-ray surveys may be missing a number of extremely obscured AGN. In the case of 
Compton-thick AGN, at column densities 
$>10^{24}$ \cunits (equivalent to $A_V\sim 450$ using the Galactic dust-to-gas ratio), a 
large fraction of the intrinsic flux will be absorbed.   
The {\it SWIFT}  (Gehrels et al. 2004) and the {\it INTEGRAL} missions \
(Winkler et al. 2003) which carry ultra-hard X-ray 
detectors ($>$15 keV), albeit  with limited imaging capabilities, 
probed energies which are  immune to X-ray obscuration up to column 
densities of $10^{25}$ \cunits. These missions helped  
towards further constraining the number density of such heavily absorbed sources  
at very bright fluxes, $\rm f_{17-60~keV} >10^{-11}$ \funits in the local Universe, $z<0.1$
(Beckmann et al. 2006, Bassani et al. 2006,  Winter et al. 2008, Winter et al. 2009).
The luminosity function in these energies has been derived by Sazonov et al. (2007),
Paltani et al. (2008) and Tueller et al. (2009).       
It appears that the fraction of Compton-thick sources is small and thus 
the {\it RXTE} and   {\it HEAO-1} luminosity functions are little  affected.

In the optical, QSOs have been traditionally selected using colours (Schmidt \& Green 1983, Marshall et al. 1987, 
Boyle et al. 2000). The advent of the 2dF (Croom et al. 2004) and the SDSS (York et al. 2000) surveys
have generated vast samples of QSOs providing a leap forward in the study of their 
luminosity function. The derived sky density of QSOs (few hundred per square degree) 
is at least an order of magnitude lower than that derived from X-ray surveys. 
This is because the colour selection of AGN requires that the nuclear optical luminosity is much higher 
than that of the host galaxy for the AGN to be detected (typically $M_B<-23$). 
Therefore the optical selection based on colours is biased against low luminosity AGN 
in contrast to the X-ray selection. 

The limitations of colour optical selection techniques can be circumvented 
 by selecting AGN via their emission lines. Such methods  
 can extend the optical luminosity function 
to low luminosities. Ho et al. (1997) carried out a spectroscopic survey of about 500 
nearby galaxies selected from the revised Shapley-Ames catalogue (Sandage \& Tammann 1981).   
They identified Seyfert emission-line characteristics in 52 galaxies.
Ulvestad \& Ho (2001) derived the optical B-band luminosity function from their sample
extending the AGN luminosity function to $\rm M_B\sim -17$ (see also Georgantopoulos et al. 1999).
A limitation in the above works is that the B-band is strongly affected by the host galaxy light.    
Hao et al. (2005a) extended significantly these results using the SDSS survey to select a sample of about 3000 AGN 
in the redshift range $0<z<0.15$.   Hao et al. (2005b) derive the $Ha$ and $[OIII]$ 
$\lambda5007$ luminosity function separately for Seyfert-1 and Seyfert-2 galaxies.
The [OIII] line is considered a much better proxy of the nuclear power as compared to the B-band luminosity,
although in the case of  strong star-forming emission some contamination of the [OIII] emission is expected.

The question which arises is whether the optical emission lines or the X-ray  selection methods are more efficient 
for finding low-luminosity AGN. This can be addressed by  comparing the X-ray and optical AGN luminosity
functions.   Heckman et al. (2005) addressed  this issue by converting the 
optical [OIII] Seyfert (combined type 1 and type 2) luminosity function of Hao et al. (2005b) to X-ray wavelengths using the 
relation between X-ray and [OIII] luminosity. Comparison with the {\it RXTE} X-ray luminosity function reveals 
that the latter  lies consistently below the optical one. As these authors have not taken into account, intentionally,  
the X-ray absorption, they conclude that the mismatch between the two 
luminosity functions can be more naturally  attributed to absorption in X-ray wavelengths. 
The overall conclusion from this work is that the optical [OIII] emission may provide a 
more efficient method for picking AGN.
In this paper we attempt to  further address this issue and to understand the 
reason for the disagreement found by Heckman et al. (2005). 
We derive the X-ray luminosity function again from the [OIII] luminosity function  of 
Hao et al. by combining it with the $\rm L_X-L_{[OIII]}$  relation  derived from 
{\it XMM-Newton} observations (Akylas \& Georgantopoulos 2009) of all the Seyferts in 
the Ho et al. sample.   Our analysis involves the following improvements:
\begin{itemize}
\item{The available  {\it XMM-Newton} spectra allow for correcting for the X-ray absorption}.
\item{We produce individually the luminosity function of Seyfert-1 and Seyfert-2 galaxies}. 
\item{We derive the X-ray luminosity function from the bi-variate optical/X-ray 
luminosity function taking fully into consideration the $\rm L_X-L_{[OIII]}$ relation and its 
dispersion}.                   
\end{itemize}             
We adopt $\rm H_o=  75  \,  km  \,  s^{-1}  \,  Mpc^{-1}$,  $\rm \Omega_{M}  =  0.3$,
$\Omega_\Lambda = 0.7$ throughout the paper.

\begin{table*}
\begin{center}
\caption{LogX-ray vs Log[OIII] luminosity relation in optically selected AGN}
\label{lxlo_ho}
\begin{tabular}{lccccc}
\hline
Sample & Number & Slope &  Intercept & $\rm <log(L_x/L_{[OIII]})>$ & Dispersion  \\
  \hline 
A. Seyfert-1 (Ho \& Heckman)& 28 & $0.84\pm0.09$  & $8.27\pm 3.87$ & 1.66 & 0.60 \\    
B. Seyfert-1 (Ho)           &  8 & $0.66\pm0.22$ & $15.31\pm8.84$ & 1.84 & 0.82 \\
C. Seyfert-1 (Heckman)      & 20 & $1.04\pm0.17$ & $-0.17\pm7.34$ & 1.59 & 0.48 \\
D. Seyfert-2  (Ho luminous+weak)      & 23 & $0.83\pm0.15$ & $7.3\pm6.1$ & 0.87 & 0.80 \\
E. Seyfert-2 (Ho luminous)  & 11  & $0.64\pm0.17$ & $15.34\pm7.15$ & 1.16 & 0.77 \\
F. Seyfert-2 (Ho weak)      & 12 & $0.59\pm0.29$ & $16.18\pm 11.48$ & 0.61 & 0.78 \\
G. Seyfert 1+ 2 (luminous)     & 17 & $0.65\pm0.14$ & $15.35\pm5.90$ & 1.44 & 0.84 \\  
\hline 
\end{tabular}
\end{center}
\end{table*}

\section{The relation between X-ray and [OIII] luminosity}
Here, we explore the relation between the X-ray and [OIII] luminosity for the optically selected Seyfert galaxies in the 
Palomar spectroscopic survey of nearby galaxies
(Ho, Filippenko, \& Sargent 1995). This survey has taken
high quality spectra of 486 bright ($B_T <$ 12.5 mag), northern
($\delta > 0^\circ$) galaxies selected from the Revised Shapley-
Ames Catalogue of Bright Galaxies (RSAC, Sandage \&
Tammann 1979) and produced a comprehensive and homogeneous
catalogue of nearby Seyfert galaxies.
Akylas \& Georgantopoulos (2009) present \xmm X-ray spectra 
for all  sources with  a reliable Seyfert classification. 
There are 38 sources in the sample of which 30 are classified 
as type-2 and 8  as type-1.  
We correct the X-ray luminosities for absorption using 
the  observed column densities given in that paper. 
There are 5 Seyfert-2 galaxies which do not show 
absorption in their individual spectra. However, their 
stacked  spectrum shows evidence for absorption. 
We choose to discard these galaxies from further analysis
as we cannot correct accurately for their absorbed luminosity. 
 We further discard two objects as they either have uncertain [OIII] flux 
  or their AGN classification has been questioned (see Akylas \& Georgantopoulos 2009).  
This leaves us with a sample of 23 Seyfert-2 galaxies. 
A large number (12) of these show no intrinsic absorption 
but still their {\lxlo} ratio is substantially lower than 
that of the absorbed Seyfert-2. 
Hereafter, we call this sub-sample the 'X-ray weak' Seyfert-2 sample,
while the remaining Seyfert-2 galaxies form the 'X-ray luminous' sample. 
 All the 11 objects in the latter sample have very high obscuring column densities ($>10^{23}$ \cunits),
  with the exception of NGC4395 which has $N_H\approx 2\times 10^{22}$ \cunits.   
As we have only a limited number of Seyfert-1 galaxies, we 
expand our sample using the 20 Seyfert-1 
galaxies in the optically selected sample of Heckman et al. which is compiled from the literature 
 (Whittle 1992; Xu, Livio \& Baum 1999).   
The X-ray luminosity of the Seyfert-1 sample of Heckman et al. (2005) is not corrected for 
absorption as the column densities in these sources are expected to be negligible. 
  The [OIII] fluxes have also not been corrected for reddening.             
             
\begin{figure}
   \begin{center}
\includegraphics[height=7.cm]{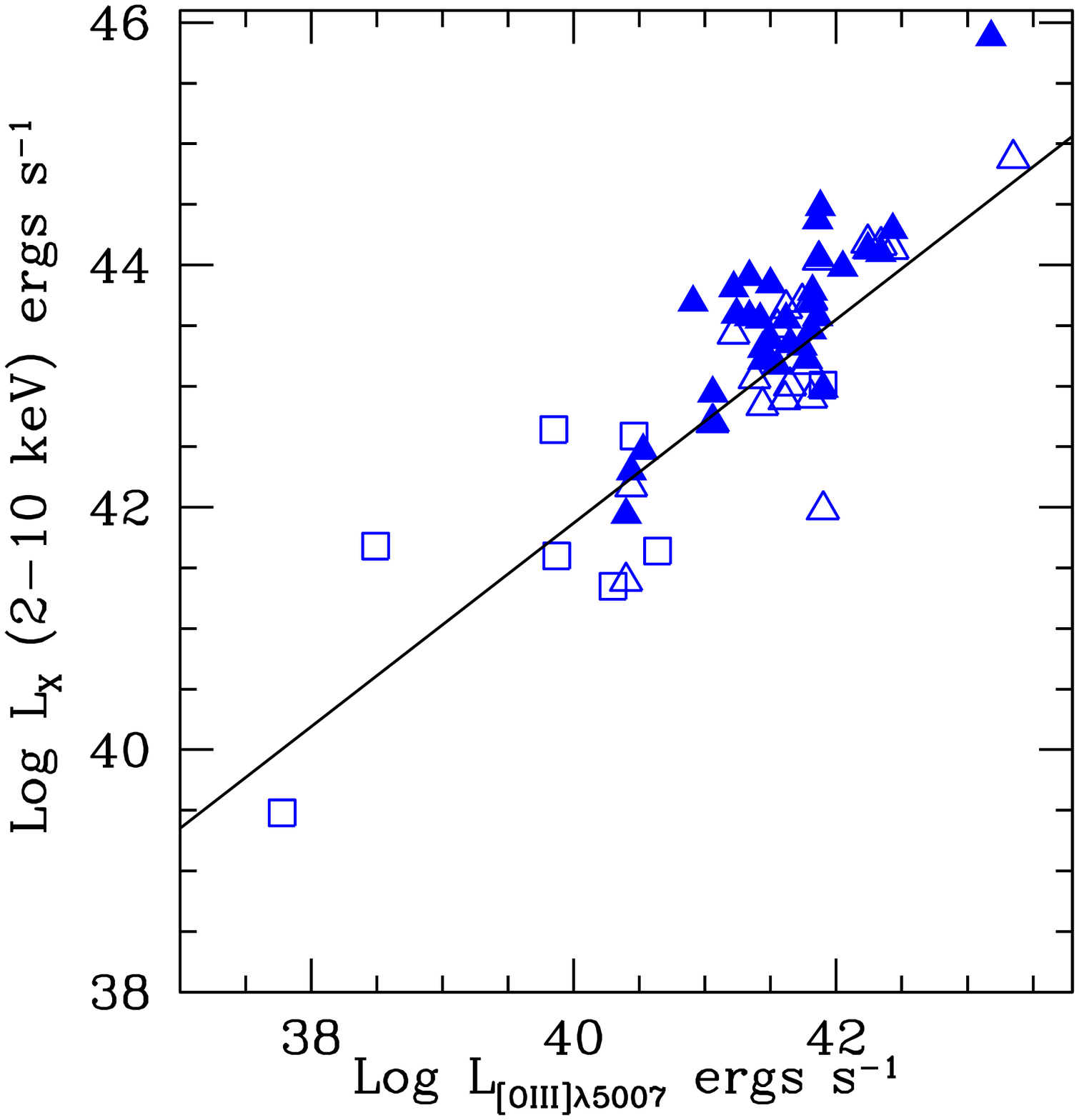}
\includegraphics[height=7.cm]{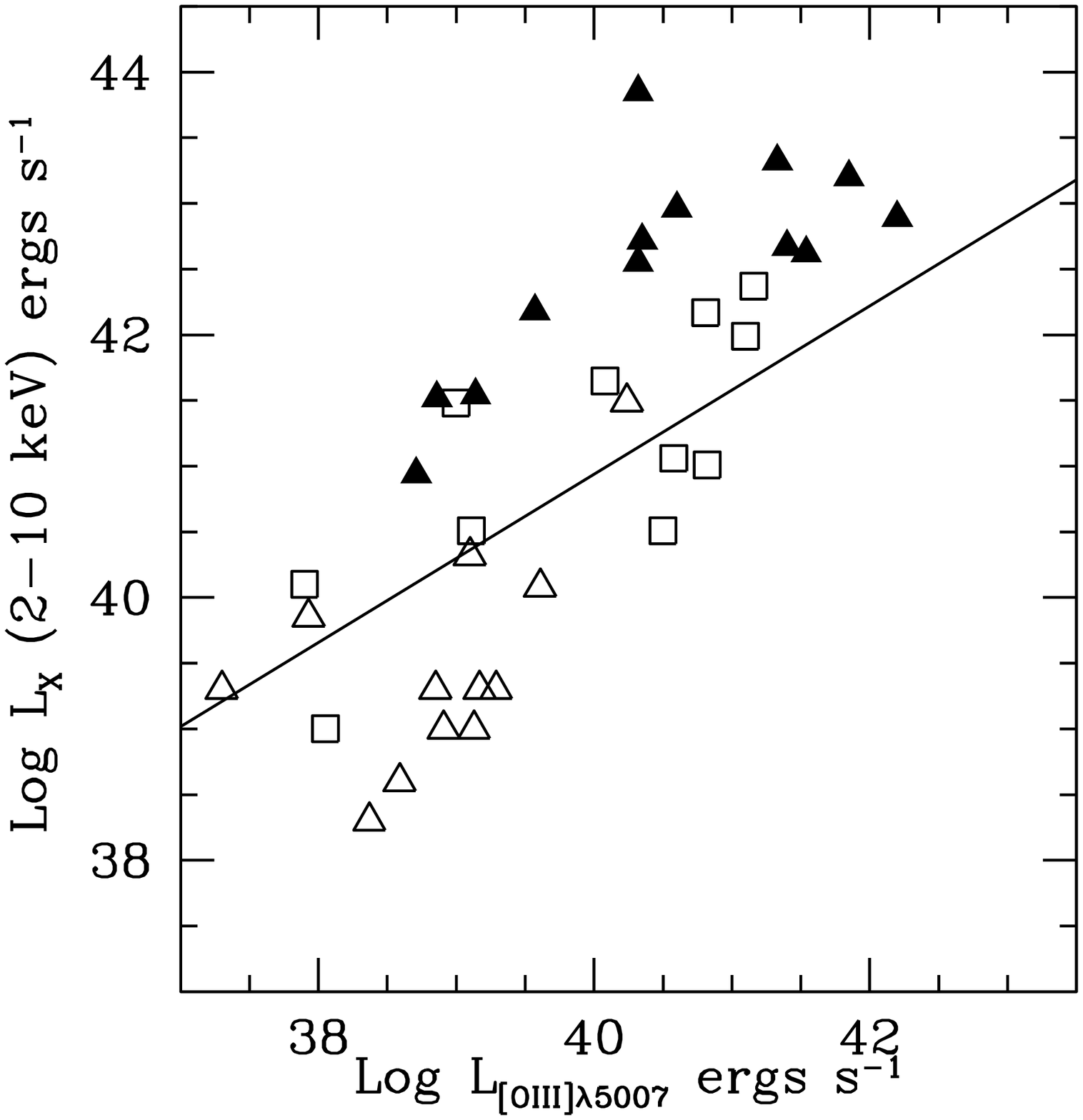}
\caption{{\bf Upper panel. Seyfert-1}: Open squares and open triangles denote sources from 
the samples of Ho et al. (sample B) and Heckman et al.  (sample C) respectively. 
Filled triangles denote the data from the sample of Sazonov \& Revnivtsev (2004). 
The solid line is the best fit result to the joint Heckman + Ho sample (sample A).   
{\bf Lower panel. Seyfert-2:} The open triangles refer to the 'X-ray weak' (unabsorbed)  
Seyfert-2 (sample F), while the open squares refer to the X-ray luminous Seyfert-2s (sample E). 
The filled triangles denote the X-ray selected sample of Sazonov \& Revnivtsev (2004).
The solid line denotes the fit to the X-ray luminous Seyfert-2  (sample E).}
\label{lxlo}
\end{center}
\end{figure}

We  fit a linear model to the $\rm \log L_X-\log L_{[OIII]} $ relation assuming no errors
in $\rm L_X$. The least square  fits are given in Table \ref{lxlo_ho}.
The {\lxlo} relation separately for type-1 and type-2 sources is shown in Fig. \ref{lxlo}. 
The best fit slope is below unity in all cases. This implies that the most luminous AGN 
have relatively less X-ray emission. 
Panessa et al. (2006) have also derived the {\lxlo} relation 
using a sample of 47 optically selected nearby AGN. 
These authors find a steeper than linear 
relation with a slope of $1.22\pm0.12$.
Their derived slopes for the type-1 and type-2 slopes are 
$0.74\pm 0.21$ and $1.28\pm 0.14$ respectively.  
However, the comparison with their sample  may not be straightforward as it 
includes  LINERS and transient objects.

Finally, for comparison, we quote the {\lxlo} relation  for the X-ray selected 
{\it RXTE} sample of Sazonov \& Revnivtsev (2004). 
This sample comprises of 76 Seyferts detected in the 3-20 keV band down to a flux limit  
of $\sim 10^{-11}$ \funits.    
The    [OIII] luminosities have been compiled by Heckman et al. (2005) for this sample.
The X-ray luminosities have been converted to the 2-10 keV band using a photon index 
of $\Gamma=1.65$  (Sazonov et al. 2008) with no correction for absorption.  
The least square fits are given in Table \ref{lxlo_saz}.
Interestingly, the Seyfert-1 follow a linear relation while the Seyfert-2 have a slope 
well below unity. The differences in the {\lxlo} relation between the 
various samples certainly  reflects the uncertainties involved 
in the determination of the predicted X-ray luminosity function. 
  
 \begin{table*}
\begin{center}
\caption{Log X-ray vs Log [OIII] luminosity relation in {\it RXTE} X-ray selected AGN}
\label{lxlo_saz}
\begin{tabular}{lccccc}
\hline
Sample & Number & Slope &  Intercept & $\rm <log(L_x/L_{[OIII]})>$ & Dispersion  \\
  \hline 
Seyfert-1    & 34  & $1.03\pm0.12$ & $0.72\pm 5.02$ & 1.98 & 0.39 \\    
Seyfert-2    & 13  & $0.52\pm0.14$ & $21.41\pm5.74$ & 2.06 & 0.77 \\
All Seyfert  & 47  & $0.88\pm0.04$ & $ 2.96\pm3.78$ & 2.00 & 0.51 \\
\hline 
\end{tabular}
\end{center}
\end{table*}

\begin{figure*}
\begin{center}
\includegraphics[height=8.5cm]{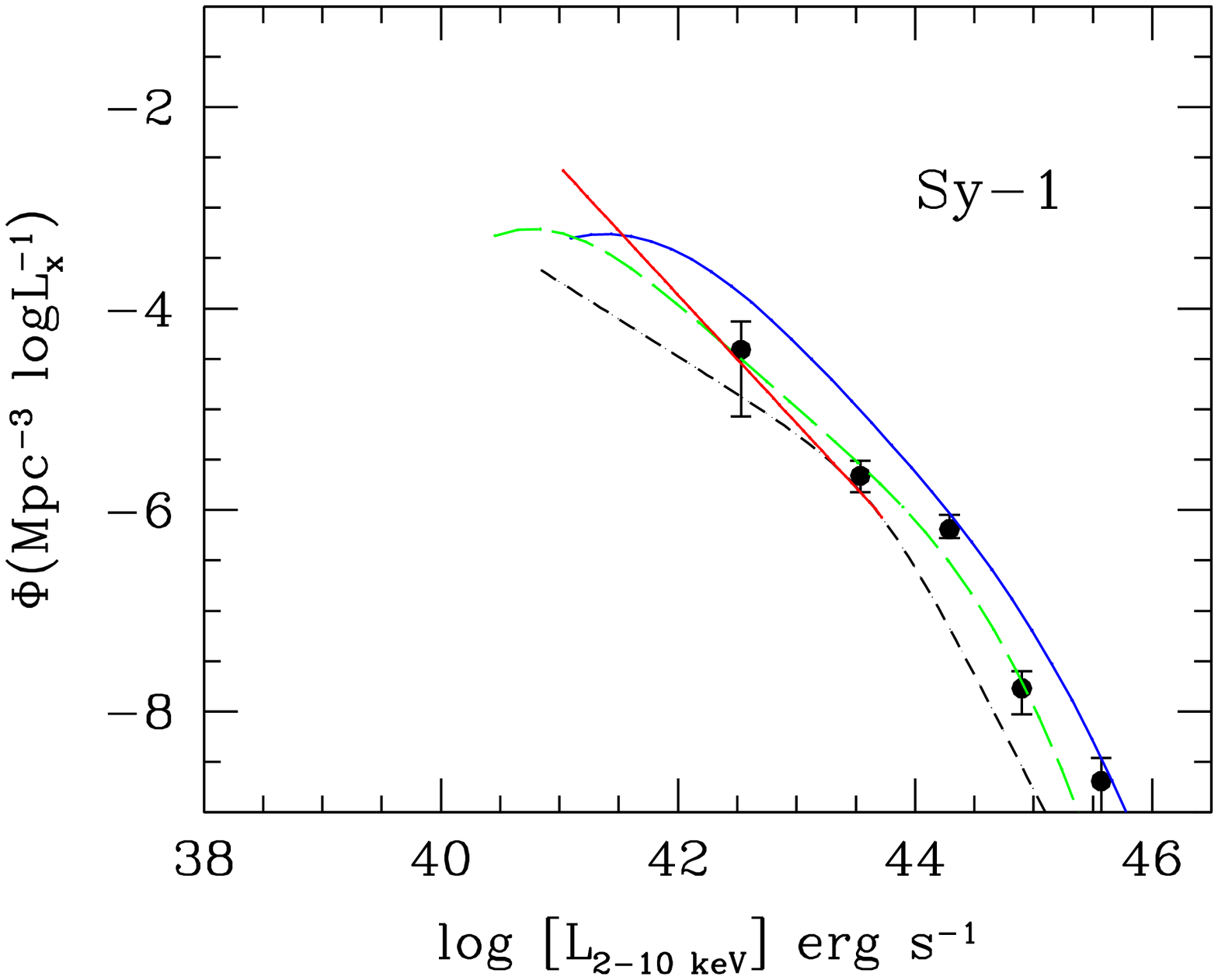}  \includegraphics[height=8.5cm]{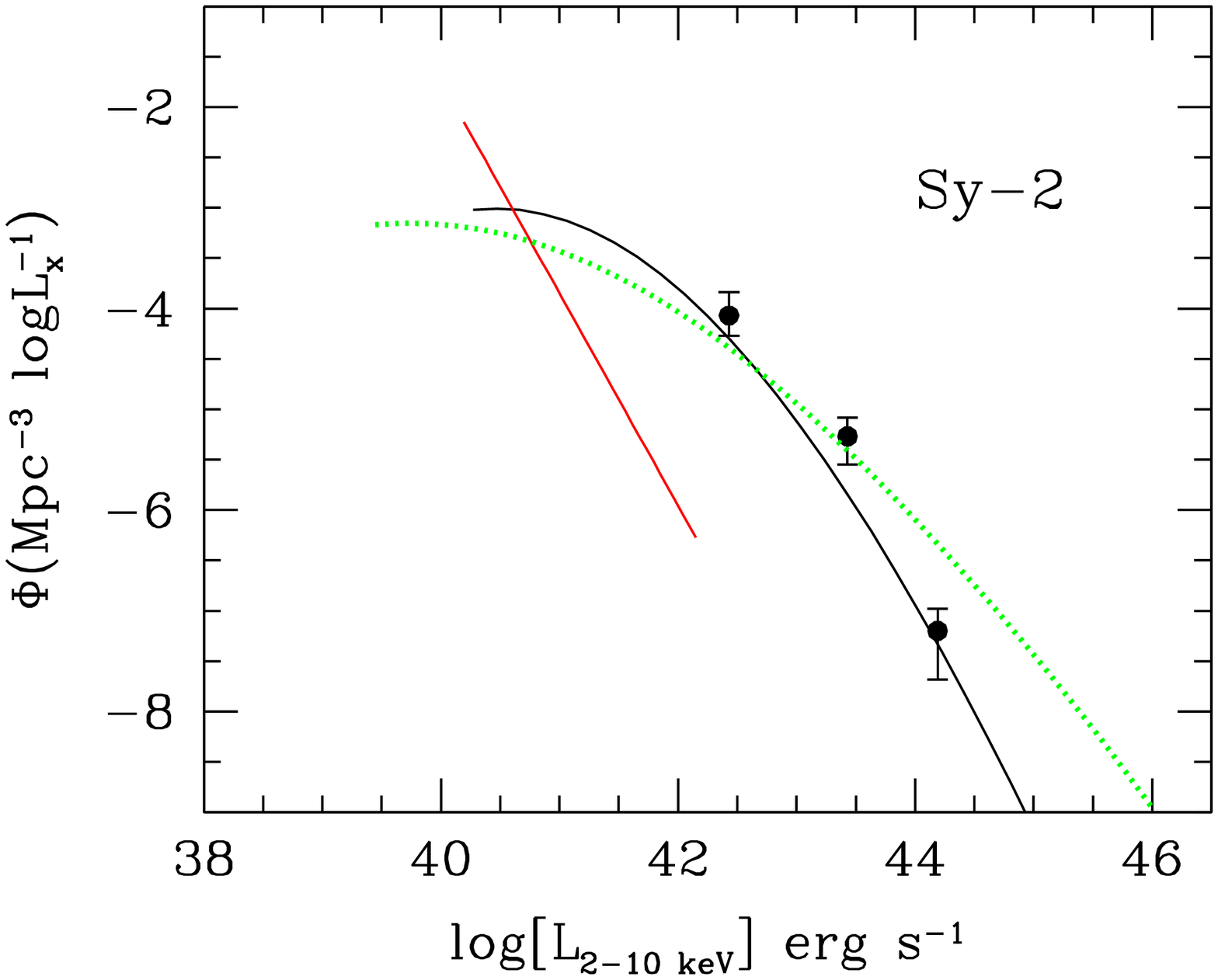}
\includegraphics[height=9.0cm]{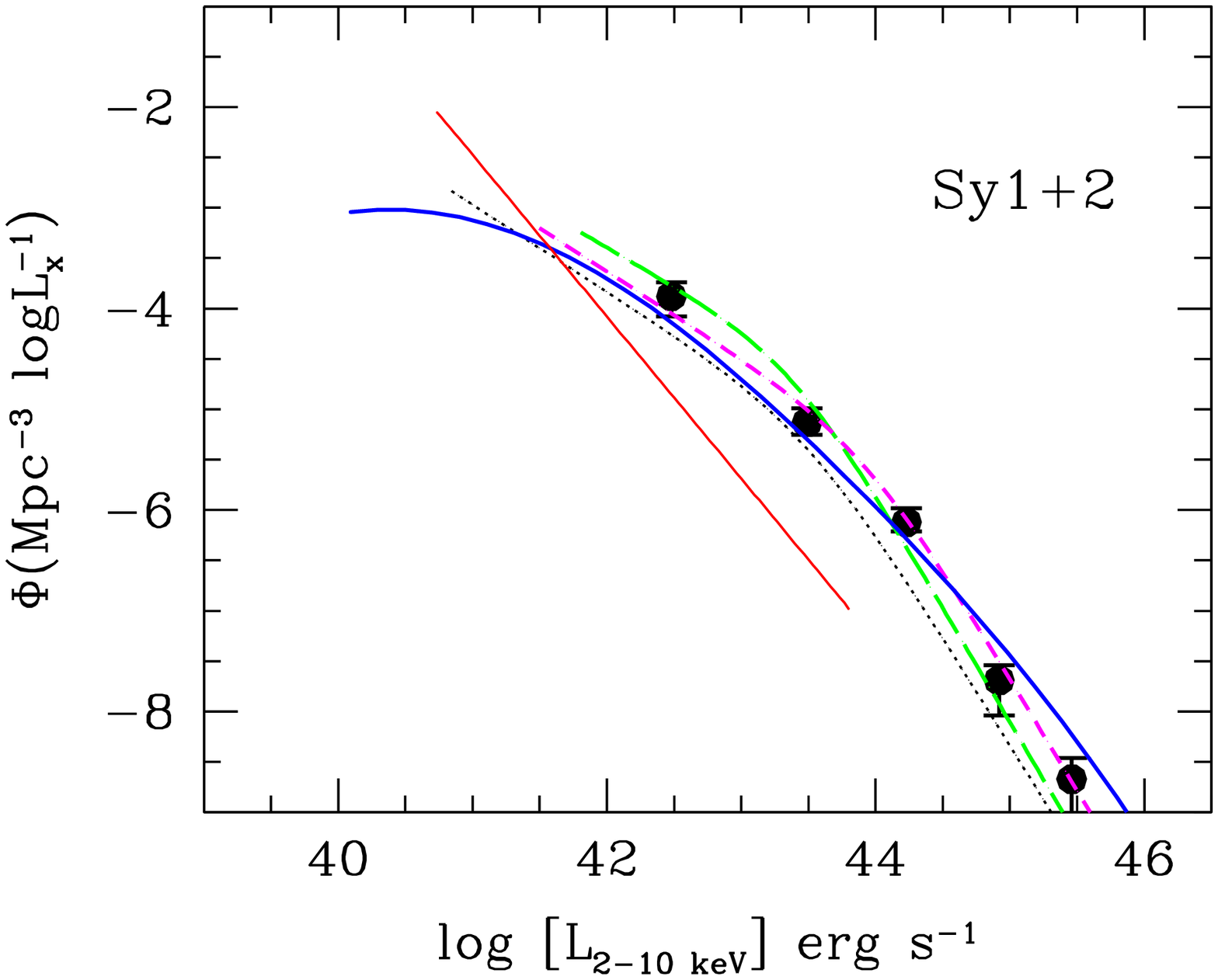}
\caption{{\bf Upper panel. Seyfert-1}: the black points denote the HEAO-1 
X-ray luminosity function of Shinozaki et al. (2006); the black (short-dash) line denotes the {\it RXTE} 
luminosity function of Sazonov \& Revnitsev (2004); the blue (solid) line and green (long-dash) lines correspond to the predicted X-ray 
luminosity function for {\lxlo} models A and C respectively. The red (solid) straight lines in all panels 
 denote the luminosity function in the case of no dispersion in the \lxlo relation ($\sigma=0$).  
{\bf  Seyfert-2: }  the black points denote the {\it HEAO-1} 
X-ray luminosity function of Shinozaki et al. (2006); the red (solid) line denotes the 
predicted luminosity function for model E while the green (dot) line denotes the 
 prediction based on the Heckman results i.e. $\rm <log(L_x/L_{[OIII]})>=0.57$  and a dispersion of $\sigma=1.06$ .
   {\bf Lower panel:  Seyfert1+2}.
The black points denote the {\it HEAO-1} luminosity function while the black (dot) line 
denotes the {\it RXTE} luminosity function for all Seyfert galaxies; the blue (solid) line corresponds to the 
predicted luminosity function (model G); the green (long-dash) and purple 
 (short-dash) curves correspond to the {\it INTEGRAL} and {\it Chandra} luminosity functions (see text for details).}
  \label{lf}
  \end{center}
\end{figure*}   

\section{Comparison between the X-ray and the optical luminosity function}

\subsection{The [OIII] luminosity function}
Hao et al. (2005a) have derived an AGN sample using 
spectroscopic SDSS data. They are using a low redshift sample
($z<0.33$) complete down to r=17.77 mag over 1151 $\rm deg^2$. 
They identify broad-line AGN 
as those having    $\rm FWHM(H\alpha)>1200 ~km~s^{-1}$.
The narrow-line AGN are selected through emission-line ratio diagnostic diagrams.
Hao et al.  (2005a) used both the diagnostic diagrams of Kauffman et al. (2003) 
and those of Kewley et al. (2001). The former criteria produce a sample which is three 
times larger but it is more prone to contamination from other low-luminosity AGN such as transient objects. 
The broad-line (Seyfert-1) and narrow-line (Seyfert-2) sample 
consist of 1317 and 3074 sources respectively.
Their [OIII] luminosities span about 3 orders of magnitude ($10^{5}-4\times 10^{8}$ $\rm L_\odot$). 
Hao et al. (2005b) have parametrised the [OIII] luminosity function 
with a Schechter, a double power-law as well as a single power-law function. 
The single power-law function  provides a very good fit to the data.
For simplicity we use this functional form thereafter.    
 We note that recently, Reyes et al. (2008) extended this work to higher 
  luminosities by constructing 
a catalog of 887 type-2 quasars from the SDSS. 
Their  luminosity function at z$<0.3$ probes luminosities 
up to $\rm 10^{9} L_\odot$ agreeing with the extrapolation of
the Hao et al. (2005b) luminosity function.

\subsection{The predicted X-ray luminosity function}  
Having a relation between the X-ray and the optical luminosity, as well as 
a functional form for the optical luminosity function, we can 
construct the bi-variate optical/X-ray luminosity function (e.g. Georgantopoulos et al. 1999). 
This is given by the convolution of the optical luminosity function with 
a probability function which defines the value of the X-ray luminosity 
at a given optical luminosity.     
Then the X-ray luminosity function is the integral of the bi-variate 
luminosity function over optical luminosity.
   
   \begin{equation} 
   \Phi(L_X)= \int \Phi(L_o)  (1/\sigma\sqrt{(2\pi)})  e^{-(L_X-<L_X>)^2/2\sigma^2}  dL_o
   \end{equation}

where $<L_X>= \alpha L_o + \beta$ with $\alpha$ and $\beta$ given in table 1.  
 The dispersion $\sigma$ plays a crucial role in the determination of the luminosity function. 
Large values of $\sigma$ result in higher numbers of X-ray sources in the 
bright X-ray luminosity bins. For  comparative reasons, we derive as well   
   the predicted X-ray luminosity function for the extreme case of no dispersion i.e. $\sigma=0$.  
The predicted X-ray luminosity functions have been constructed as follows. 

\begin{itemize}
{\item Seyfert-1. We use the optical Seyfert-1 luminosity function of Hao et al.
 in combination with  the {\lxlo} relation (samples A and C) in table 1.}
{\item Seyfert-2. We use the Seyfert-2 optical luminosity function  of Hao et al.
  which is based on the Kewley criteria. 
 We  use the {\lxlo} relation based on sample E (i.e.  
excluding the X-ray weak Seyferts). Moreover, for comparison we use the 
 (not corrected for X-ray absorption) relation of Heckman et al. (2005) i.e.  $\rm <log(L_x/L_{[OIII]})>=0.57$  with a dispersion of $\sigma=1.06$ }
{\item  Seyfert1+2. 
For the total sample we use the Seyfert luminosity function of Hao et al. 
 combined with the {\lxlo} 
relation derived from sample G (see Table 1).}
 \end{itemize}
      
\subsection{Comparison with the observed X-ray luminosity function} 
The X-ray luminosity has been derived in the local Universe using the {\it RXTE} 
slew survey data (Sazonov \& Revnivtsev 2004). These authors construct a sample 
of 76 non-blazar AGN, with luminosity $\rm L_x> 10^{41}$ \lunits,  
most of them populating low redshifts $z<0.1$. 
They derive the luminosity function for the total sample  as well as separately for the 
unabsorbed sample having column densities $\rm N_H<10^{22}$  \cunits.   
The unabsorbed sources (type-1) are 59 and form the vast majority of the sample.
We note that 60 sources are classified as Seyfert-1 
{\it on the basis of their optical classification} and thus  there is good agreement 
 between the X-ray and optical type-1 classification.  
The total luminosity function is described by a double power-law form with a break 
luminosity of $\rm L(3-20 keV) \approx 4\times 10^{43}$ \lunits.  
Shinozaki et al. (2006) have derived the luminosity function of 49 AGN 
from the HEAO-1 all-sky survey in the 2-10 keV band, 
down to  flux limit of $2.7\times 10^{-11}$ \funits. 
These authors choose to use, as a  dividing line between 
type-1 and type-2 AGN,  a column density of 
 $\rm N_H =10^{21.5}$ \cunits.  
      
In Fig.\ref{lf} we compare our predicted luminosity function with the
X-ray luminosity functions described above.  
The total predicted luminosity function (Sy1+2),  agrees quite well with with the
luminosity function of Shinozaki et al. (2006) and that of Sazonov et al. (2004)    
For comparison, we also plot the luminosity function of Ueda et al. (2003) 
at z=0 as well as the {\it INTEGRAL} luminosity function derived by Sazonov et al. (2007). 
in the 17-60 keV band. For the conversion to the 2-10 keV band luminosity    
we assumed a slope of $\Gamma=1.65$ (Sazonov et al. 2008).   
 It is instructive to examine the relative fraction of Seyfert-1 and Seyfert-2 
in the optical samples  as compared with that in the X-ray selected samples.
 In the latter, there is a preponderance of Seyfert-1
(about 3.3:1 in {\it RXTE}). 
 In contrast, integration of the optical luminosity functions of Seyfert-1 and Seyfert-2
shows that the two populations have approximately the same density. Only in the brighter regime 
($\log L_{[OIII]} = 40.2- 42.14$) where there is overlap with the X-ray luminosity function, 
the number of Seyfert-1 is 2.5 times higher than that of Seyfert-2 (rather more in line with
 the X-ray selected samples). 

Next, we compare luminosity functions for type-1 and 2 Seyferts separately. 
 We have to bear in mind that, the optical luminosity function  (and thus the predicted X-ray luminosity function) 
is based on optical line classification while the observed X-ray luminosity function 
is based on X-ray spectroscopy. Nevertheless, as we saw above
the correspondence between optical and X-ray classification is reasonably good. 
The predicted Seyfert-1 luminosity function is somewhat higher than 
the observed one, in the case of the {\lxlo} relation (sample A). 
When we use  instead the {\lxlo} relation derived from sample C
the agreement is  better.     
In the case of the  Seyfert-2 sample  the agreement between the 
 optical and X-ray luminosity function is quite reasonable.
  In the Seyfert-2 samples it is possible that a low \lxlo ratio may be suggestive of 
   excess absorption above what is directly seen in the X-ray spectra (e.g. 
    Melendez et al. 2008, LaMassa et al. 2009). This may indeed hold true 
     in some of the X-ray 'weak' Seyfert-2  (sample F). 
      However, this does not appear to be the case among the 
       'luminous' Seyfert-2 (sample E) that we are using for the derivation 
        of the Seyfert-2 luminosity function. All  these Seyferts show 
         large obscuring columns and thus the correction to the 
          intrinsic luminosity has been straightfoward.

\section{Discussion \& Summary}

We derived the predicted X-ray luminosity function for Seyfert galaxies in the local Universe,  
by combining the optical SDSS [OIII] Seyfert luminosity functions with the corresponding  {\lxlo} relation. 
These relations have been derived  using {\it XMM-Newton} observations
(Akylas \& Georgantopoulos 2009) of the local, optically selected AGN sample of Ho et al. (1997). 
This sample covers a comparable luminosity range with the SDSS Seyfert sample.   
We have corrected the X-ray luminosity for the effects of absorption.  
Our analysis above shows that the predicted X-ray luminosity function 
is in reasonable agreement with the observed X-ray  Seyfert luminosity functions
derived in the 2-10 keV and 2-20 keV bands by {\it HEAO-1} and {\it RXTE} respectively.
Most importantly, it is in agreement with the  ultra-hard 17-60 keV {\it INTEGRAL}
luminosity function (Sazonov et al. 2007). 
 As the {\it INTEGRAL} luminosity function is practically immune to X-ray absorption, 
  this suggests that absorption played little role 
in the optical/X-ray luminosity function discrepancy reported by Heckman et al. (2005).    
This is also independently supported by the {\it XMM-Newton} observations of the 
Palomar optically selected Seyfert sample (Akylas \& Georgantopoulos 2009). 
 The fraction of Compton-thick AGN in this optically selected sample is small.  
     
In addition we examined separately,  
the Seyfert-1 and Seyfert-2 luminosity function.
The Seyfert-1 luminosity function is in rough agreement or even somewhat above
the observed X-ray luminosity function.   
The predicted Seyfert-2 luminosity function agrees quite well with the 
optical luminosity function again disfavouring the absorption hypothesis.   
Indeed, if absorption were the problem, then our predicted luminosity function, 
which takes X-ray absorption into account, would be well above that of {\it RXTE}. 
    
Having addressed this matter, we need to understand  why  
Heckman et al. (2005) found that the {\it RXTE} X-ray luminosity function 
 lies below the optical one by about a factor of 3. 
The comparison was based on the fact that the two luminosity functions have the same slope. 
  However,  this is true only  
for the  bright part of the X-ray luminosity function 
(i.e. for L$_{\star}(3-20~ \rm keV)>3\times 10^{43}$ \lunits or equivalently 
L$_{\star}(2-10 \rm keV)>2.3 \times 10^{43}$ \lunits). Using  
$\rm (L_X/L_{[OIII]}) \approx 100 $ the corresponding optical 
luminosities must be  greater than $5.7 \times 10^7 L_{\odot}$. 
Since the upper end of the optical luminosity function of Hao et al. (2005b) 
 is $\sim 3 \times 10^8 L_{\odot}$  the  overlap is very limited (see Fig.5 in Heckman et al.) 
  and thus the comparison is not quite robust. Moreover, the effect of the dispersion 
   plays a critical role as shown here.  

Although, the densities of the [OIII] and the X-ray selected AGN  are comparable,
this does not mean that  these methods favour the selection of the same objects. 
The optical selection favours X-ray weak Seyfert-2 (see Fig. 1). 
It has been proposed (see Ho 2008 for a review) that a large number of 
sources at low luminosities appear as Seyfert-2 possibly 
because no  Broad-Line-Region is formed at 
low accretion rates (Nicastro 2000) or low luminosities 
(Elitzur  \& Shlosman 2006).  If the above models hold true, 
these 'naked' Seyfert-2 galaxies should present no hidden Broad Lines 
in spectropolarimetric observations (e.g. Tran 2003).
Interestingly, Spinoglio et al. (2009) find that the  ratio of the X-ray to 
12$\mu m$ luminosity, $L_X/L_{12\mu m}$, 
is much lower in the Seyfert-2 with no hidden Broad-Line-Region,
suggesting that these are weak X-ray emitters. 
There is however, one caveat in these 'naked' X-ray weak 
Seyfert-2 interpretation. The [OIII] (or the mid-IR) luminosity
is believed to be a good proxy of the ionizing nuclear 
luminosity and thus to the X-ray luminosity.
But in these Seyfert-2 sources the $L_X/L_{[OIII]}$ ratio 
is low, implying that only the X-ray luminosity is weak.
One explanation could be that star-formation is 
contributing a large part of the  [OIII] emission.      
   
The future missions {\it NUSTAR} and {\it ASTRO-H} having superb imaging 
capabilities ($\sim$ 1arcmin) at ultra-hard energies (10-70 keV),
will provide the opportunity to obtain the least unbiased 
AGN samples, reaching  flux levels at least two orders of magnitude fainter    
than {\it SWIFT} and {\it INTEGRAL}.

\end{document}